\documentclass[aps,prl,twocolumn]{revtex4-1}

\usepackage[utf8]{inputenc}
\usepackage{amsmath}
\usepackage{graphicx}
\usepackage{bbold}
\usepackage{soul}
\usepackage{xcolor}
\usepackage{tabularx}
\usepackage{multirow}
\usepackage{makecell}

\DeclareMathOperator{\Tr}{Tr}

\begin{document}

\title{Emergent correlated phases in rhombohedral trilayer graphene induced by proximity spin-orbit and exchange coupling}

\author{\surname{Yaroslav} Zhumagulov$^{1}$}
\email[Emails to: ]{iaroslav.zhumagulov@ur.de}
\author{\surname{Denis} Kochan$^{2,1}$}
\author{\surname{Jaroslav} Fabian$^{1}$}
\email[Emails to: ]{jaroslav.fabian@ur.de}

\affiliation{%
 1) Institute for Theoretical Physics, University of Regensburg,
 93040 Regensburg, Germany\\
 2) Institute of Physics, Slovak Academy of Sciences, 84511 Bratislava, Slovakia
 }%

\date{\today}

\begin{abstract}
The impact of proximity-induced spin-orbit and exchange coupling on the correlated phase diagram of rhombohedral trilayer graphene (RTG) is investigated theoretically. By employing \emph{ab initio}-fitted effective models of RTG encapsulated by transition metal dichalcogenides (spin-orbit proximity effect) and ferromagnetic Cr$_2$Ge$_2$Te$_6$ (exchange proximity effect), we incorporate the Coulomb interactions within the random-phase approximation to explore potential correlated phases at different displacement field and doping. We find a rich spectrum of spin-valley resolved Stoner and intervalley coherence instabilities induced by the spin-orbit proximity effects, such as the emergence of a \textit{spin-valley-coherent} phase due to the presence of valley-Zeeman coupling. Similarly, proximity exchange removes the phase degeneracies by biasing the spin direction, enabling a magneto-correlation effect---strong sensitivity of the correlated phases to the relative magnetization orientations (parallel or antiparallel) of the encapsulating ferromagnetic layers. 
\end{abstract}

\maketitle

\paragraph{Introduction.} The discovery of correlated phases and superconductivity in magic-angle twisted bilayer graphene~\cite{Dean2013, Kim2016, Cao2018_1, Cao2018_2}, which exhibits a highly flat band structure at the
Fermi level~\cite{Li2009, Bistritzer2011, PhysRevLett.109.196802, Ponomarenko2013, PhysRevB.82.121407, TramblydeLaissardire2010,PhysRevB.81.165105, PhysRevLett.106.126802, PhysRevB.85.195458, PhysRevB.86.155449, PhysRevLett.99.256802},
has prompted intense theoretical~\cite{PhysRevResearch.3.033260, PhysRevB.98.045103, PhysRevResearch.2.033150, PhysRevB.101.125411, PhysRevX.8.031089, PhysRevLett.124.097601, PhysRevB.95.075420, PhysRevLett.127.147203, PhysRevB.98.241407, PhysRevB.98.245103, PhysRevB.99.165112} and experimental~\cite{Yankowitz2019, Lu2019, Xie2021, PhysRevLett.123.197702, Rickhaus2021,Choi2019,Liu2020, Wong2020, Zondiner2020, Kerelsky2019,Jiang2019, Polshyn2019, PhysRevLett.128.217701,Codecido2019, Balents2020} investigations. 
However, while twist angle has emerged as a new tuning knob of the electronic properties of van der
Waals heterostructures, it remains a challenge to control the stacking angle and limit twist disorder~\cite{Beechem2014, Uri2020, Gadelha2021, PhysRevResearch.2.023325, Kazmierczak2021, PhysRevB.105.245408, PhysRevB.102.064501,PhysRevB.107.L081403}. 

As shown recently, correlation phenomena are not exclusive to moir\'e structures. Observations of half and quarter metallic states~\cite{Zhou2021_1, PhysRevB.105.L081407, PhysRevB.106.155115, PhysRevB.107.L121405, arxiv.2305.04950, Geisenhof2021} and superconductivity~\cite{Zhou2021_2, PhysRevLett.127.187001, PhysRevLett.127.247001, PhysRevB.105.134524, Chatterjee2022, PhysRevB.106.155115, PhysRevB.105.075432, arxiv.2203.09083, PhysRevB.107.L161106, Pantaleón2023, li2023charge} in rhombohedral trilayer graphene (RTG), and
of isospin magnetism and spin-polarized superconductivity in Bernal bilayer graphene~\cite{Zhou2022, Zhang2023, Seiler2022, PhysRevB.105.L100503, delaBarrera2022, arxiv.2303.00742, PhysRevB.105.L201107, PhysRevB.106.L180502, PhysRevB.107.L161106, Pantaleón2023, li2023charge}, have demonstrated that 
rich physics of strong electronic correlations can be manifested in more conventional, moir\'e-less graphene systems. The key feature shared by magic-angle twisted bilayer graphene with Bernal-stacked bilayer graphene and RTG, is the presence of pronounced van Hove singularities (vHS) in the electronic density of states near the charge neutrality point~\cite{Li2009, Bistritzer2011, PhysRevB.82.035409}. In RTG, the low energy of vHS allows tuning the induced symmetry-breaking phases by doping~\cite{Zhou2021_1}, while its multilayered atomic structure enables efficient control of the vHS and thus correlated phases via a displacement field~\cite{PhysRevB.82.035409, Zhou2021_1}. This tunability makes RTG a promising platform for exploring strongly correlated physics.

\begin{figure}
    \centering
    \includegraphics[width=.5\textwidth]{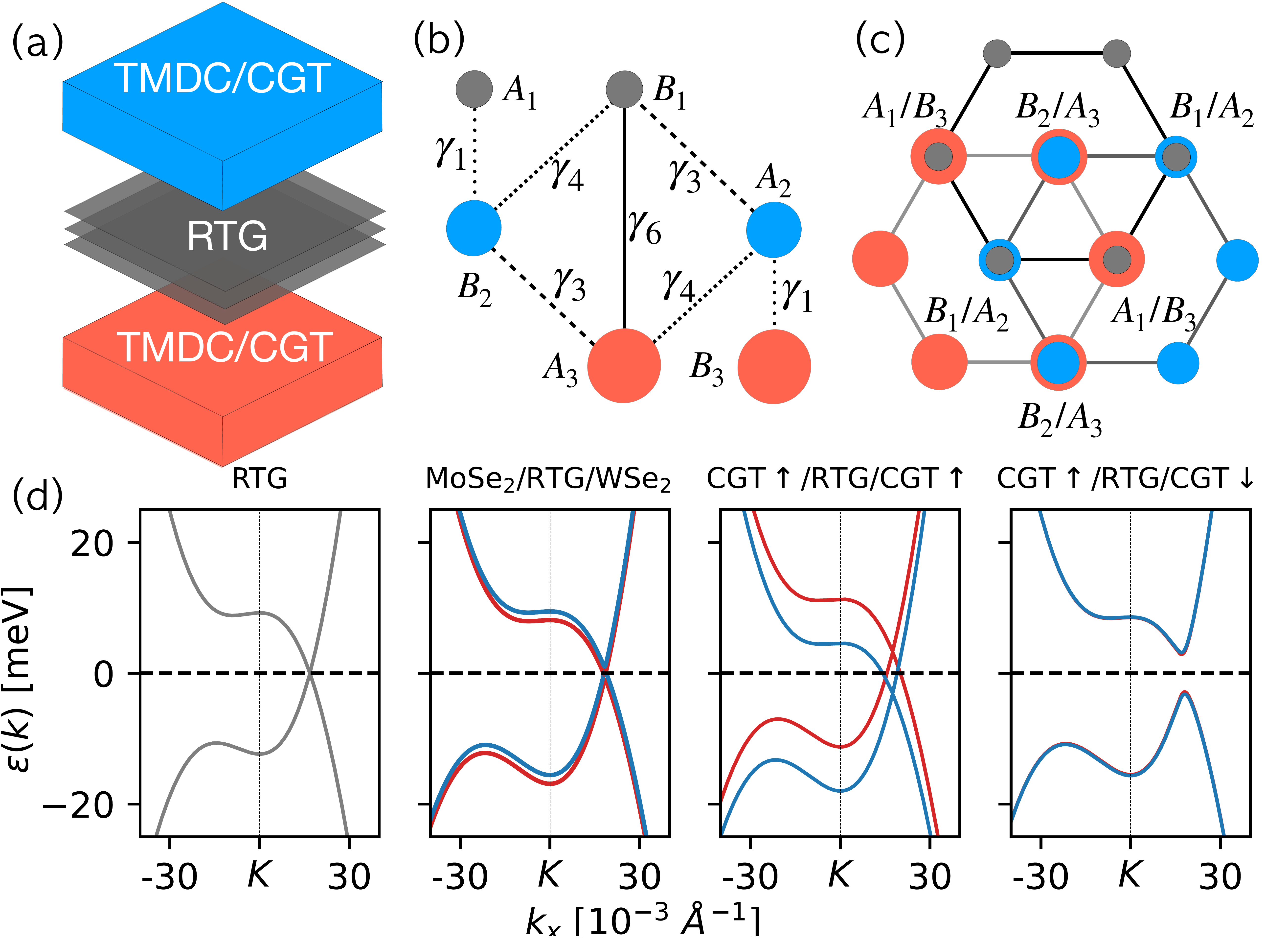}
    \caption{
    (a) Scheme of an RTG-based heterostructure encapsulated by transition-metal dichalcogenide (TMDC) or CGT monolayers which induce, respectively, SO or EX interactions in RTG. 
    (b) RTG unit cell with the relevant interlayer orbital hoppings $\gamma$. 
    The colors distinguish the three layers and $A_l$ and $B_l$ denote sublattice sites of the $l$th layer. 
    (c) The top-down view of the RTG lattice. 
    (d) Calculated single-particle low-energy electronic dispersions
    at $K$ [from $\hat{h}({\bf k}, \tau)$ without displacement fields]  
    for pristine RTG and indicated heterostructures (red marks spin up and blue spin down states).
    }  
    \label{fig:scheme}
\end{figure}

Electronic band structures can also be tuned by van der Waals engineering. In particular, proximity-induced
spin interactions---spin-orbit (SO) and exchange (EX) couplings---can furnish graphene with what it lacks: SO fields and spin polarization. Indeed, 
proximity-induced SO and EX interactions in graphene-based heterostructures have been theoretically predicted~\cite{PhysRevB.99.085411, PhysRevB.104.195156, Gmitra2015, Gmitra2016, PhysRevB.104.075126, PhysRevLett.125.196402, Gmitra2013, PhysRevB.105.115126, PhysRevB.94.155441, PhysRevB.77.115406, Kochan2017, PhysRevB.104.075126, PhysRevLett.125.196402, PhysRevB.105.115126, ZollnerTwist} and experimentally observed~\cite{Garcia2018, Island2019, Hoque2021, Ghiasi2017, Ghiasi2019, Safeer2019, Herling2020, Wakamura2019, Wakamura2020, InglaAyns2021, Kaverzin2022, Karpiak2019}, demonstrating
the appearance of valley-Zeeman, Kane-Mele, and Rashba SO coupling ~\cite{Garcia2018, Island2019, Hoque2021, Ghiasi2017, Ghiasi2019, Safeer2019, Herling2020, Wakamura2019, Wakamura2020, InglaAyns2021, Kaverzin2022, PhysRevB.99.085411,PhysRevB.104.195156, Gmitra2015, Gmitra2016, PhysRevB.104.075126, PhysRevLett.125.196402, Gmitra2013, PhysRevB.105.115126} as well as (anti)ferromagnetic EX couplings~\cite{ZollnerTwist, PhysRevB.94.155441, PhysRevB.77.115406, PhysRevB.99.195452, PhysRevB.104.075126, PhysRevLett.125.196402, PhysRevB.105.115126, Kaverzin2022,Karpiak2019, ZollnerTwist} on the meV scale. 
It should even be possible to swap SO and EX couplings via displacement field~\cite{PhysRevB.104.075126, PhysRevLett.125.196402}. 

It is natural to ask what effects can arise from combining correlated physics and proximity-induced spin interactions. Concerning RTG, two observations are crucial: (i) The vHS are formed by the bands comprising the top and bottom layer $p_z$ orbitals, and (ii)  the spin proximity effects are on the meV scale, which is also expected for the correlated band gaps. The above suggests that spin proximity effects can significantly alter symmetry-broken phases in RTG.

In this paper, we demonstrate that indeed proximity SO and EX interactions induce and control novel strongly correlated phases. Specifically, we study MoSe$_2$/RTG/WSe$_2$ heterostructures for the  SO proximity effect, and CGT/RTG/CGT heterostructures with parallel and antiparallel magnetizations of CGT (Cr$_2$Ge$_2$Te$_6$) for the EX proximity effect~\cite{PhysRevB.105.115126}, see Fig.~\ref{fig:scheme}(a). Employing the random-phase approximation (RPA)~\cite{PhysRev.82.625, PhysRev.85.338, PhysRev.92.609, PhysRevLett.101.087004, Graser2009, PhysRevB.83.100515}, we first calculate the correlated phase diagram of pristine RTG which exhibits either an intervalley coherent (IVC) state ~\cite{PhysRevLett.103.216801} or a Stoner instability, as already predicted~\cite{PhysRevB.105.134524, Chatterjee2022}.  The spin interactions remove the degeneracies of IVC and Stoner phases by introducing spin anisotropy and spin bias. Novel phases, such as a \textit{spin-valley-coherent} (SVC) state, arise, primarily due to proximity-induced valley-Zeeman coupling; Rashba coupling plays a lesser role. The SVC phase opens new perspectives  for spintronics ~\cite{RevModPhys.76.323}, since the Coulomb interactions induce novel spin-valley couplings. Finally, the magnetic heterostructures exhibit  strong magneto-correlation effect---the induced phases are sensitive to the relative orientation of the CGT magnetizations.


\paragraph{Model.} We model the orbital physics of proximitized RTG by a realistic hopping Hamiltonian~\cite{PhysRevB.82.035409, PhysRevB.105.115126, Konschuh2011, PhysRevB.80.165409, PhysRevB.87.045419}:
\begin{equation}
\centering
    \hat{h}_0\
=\begin{pmatrix}
  \Delta+u_d & \gamma_0 f&\gamma_4 f^{*}&\gamma_1&0&0 \\ 
  \gamma_0 f^{*} & \eta+u_d&\gamma_3 f&\gamma_4 f^{*}&\gamma_6&0\\
  \gamma_4 f&\gamma_3 f^{*}&\Delta+u_m&\gamma_0 f&\gamma_4 f^{*}&\gamma_1\\
  \gamma_1&\gamma_4 f&\gamma_0 f^{*}&\Delta+u_m&\gamma_3 f&\gamma_4 f^{*}\\
  0&\gamma_6&\gamma_4 f&\gamma_3 f^{*}&\eta-u_d&\gamma_0 f\\
  0&0&\gamma_1&\gamma_4 f&\gamma_0 f^{*}&\Delta-u_d
\end{pmatrix}.
\label{eq:six}
\end{equation}
acting on single-particle, $p_z$-orbital Bloch states with momenta $\mathbf{k}=(k_x, k_y)$ measured 
from $K$ and $K'$ valleys.  
Here, $f=-\left(\sqrt{3}a/2\right)\left(\tau k_x-ik_y\right)$ is the linearized nearest-neighbor structure factor, $a$ is graphene's lattice constant, and $\tau_{K/K'}=\pm1$ is the valley index; ${\gamma_i}$ are orbital hopping parameters, see Fig.~\ref{fig:scheme}~(b). 
We use the orbital basis ordered according ($A_1, B_1, A_2, B_2, A_3, B_3$), where $A_l$ and $B_l$ represent the layer-resolved sublattices, with $l = 1, 2, 3$. 
The unit cell and lattice structure of RTG are shown in Fig.~\ref{fig:scheme}~(b)~and~(c). The electrostatic potentials on different layers are incorporated into Eq.~(\ref{eq:six}) through on-site energies $u_{d}$ and $u_m$. The parameter $u_m$ represents the potential energy difference between the central layer and the average potential of the outer layers. In contrast, $u_d$ corresponds to the potential energy difference between the external layers, describing an effect of the displacement field. 
Finally, $\eta$ and $\Delta$ are on-site potentials, whose asymmetry arises due to the vertical hoppings $\gamma_1$ and $\gamma_6$, see Fig.~\ref{fig:scheme}(b).

\begin{table*}[t]
  \centering
\begin{tabular}{|c|cc|ccc|}
\hline 
pristine RTG phases & 
\multicolumn{2}{c|}{IVC}& \multicolumn{3}{c|}{Stoner} \\
\hline
proximity-furnished phases & 
\multicolumn{1}{c|}{CDW$_{\pm}$} &
\multicolumn{1}{c|}{SVC$_{\pm}$} &
\multicolumn{1}{c|}{SVP$_{\pm}$} &
\multicolumn{1}{c|}{VP$_{\pm}$} &
\multicolumn{1}{c|}{SP$^{x/y}_{\pm}$} \\
\hline 
spin-valley order operators   & 
\multicolumn{1}{c|}{
\makecell{$\hat{\Psi}^\dagger(\left[s_0 \pm s_z\right]\tau_x)\hat{\Psi}$}
}     & 
\multicolumn{1}{c|}{
\makecell{$\hat{\Psi}^\dagger(s_x\tau_x\pm s_y\tau_y)\hat{\Psi}$}
}    &  
\multicolumn{1}{c|}{
\makecell{$\hat{\Psi}^\dagger (s_z\tau_0\pm s_0\tau_z)\hat{\Psi}$}
} & 
\multicolumn{1}{c|}{
\makecell{$\hat{\Psi}^\dagger(\left[s_0 \pm s_z\right]\tau_z)\hat{\Psi}$}
} & 
\multicolumn{1}{c|}{
\makecell{$\hat{\Psi}^\dagger(s_{x/y}\left[\tau_0 \pm \tau_z\right])\hat{\Psi}$}
} \\
\hline
\end{tabular}
\caption{
List of local fermionic bilinear operators $\hat{\Phi}$ in spin-valley channels for the relevant symmetry-broken phases in RTG. Here $\hat{\Psi}=(\hat{c}_{\uparrow K}, \hat{c}_{\uparrow K^\prime}, \hat{c}_{\downarrow K}, \hat{c}_{\downarrow K^\prime})^\top$ and $\hat{\Psi}^\dagger=(\hat{\Psi})^\dagger$ stand for spin-valley-resolved degrees of freedom.
}
\label{tab}
\end{table*}

Pristine RTG exhibits weak SO~coupling at $K$ and $K'$ valleys, on the scale of
$10~\mu$eV~\cite{Konschuh2011, PhysRevB.87.045419}. Thus, we consider RTG encapsulated by strong 
SO~materials MoSe$_2$ and WSe$_2$, and by ferromagnetic CGT; as shown by \emph{ab initio} simulations 
they induce spin splittings on the 1~meV scale~\cite{PhysRevB.105.115126}.
The corresponding proximity-induced SO and EX coupling Hamiltonian $\hat{h}_{\text{prox}}=\sum_{l} \hat{h}^{l}_{R}+\hat{h}^{l}_{I}+\hat{h}^{l}_{\text{ex}}$ of RTG electrons is the sum of Rashba, $\hat{h}^{l}_{R}$, intrinsic, $\hat{h}^{l}_{I}$, and exchange, $\hat{h}^{l}_{\text{ex}}$, terms in the 
given $l$th layer~\cite{PhysRevB.105.115126, Gmitra2016,Kochan2017,Gmitra2013},
parameterized, correspondingly, by the sublattice-resolved couplings $\lambda^{l}_R$, $\lambda^{A_l/B_l}_I$ and $\lambda^{A_l/B_l}_{\text{ex}}$:
\begin{align}
    \hat{h}^{l}_{R}&=
    \begin{pmatrix}
        0&2i\lambda^{l}_R\,s_{-}^{\tau}\\
        -2i\lambda^{l}_R\,s_{+}^{\tau}&0
    \end{pmatrix},\\
    \hat{h}^{l}_{I}&= \begin{pmatrix}
        \tau\,\lambda_I^{A_l}\,s_z&0\\
        0&-\tau\,\lambda_I^{B_l}\,s_z
    \end{pmatrix},\\
    \hat{h}^{l}_{\text{ex}}&= \begin{pmatrix}
        -\lambda_{\text{ex}}^{A_l}\,s_z&0\\
        0&-\lambda_{\text{ex}}^{B_l}\,s_z
    \end{pmatrix}.
    \label{eq:prox}
\end{align}
Here $s_{x,y,z}$ are the Pauli matrices acting on spin degrees of freedom and $s_{\pm}^{\tau} = \frac{1}{2}(s_x \pm i\tau s_y)$ are the valley-resolved spin-flip operators. Each $\hat{h}^l$ is $4\times 4$ matrix in the spin-sublattice resolved $(A_{l\uparrow},A_{l\downarrow},B_{l\uparrow},B_{l\downarrow})$ Bloch basis. 
Due to the short-range nature of the proximity effects, we consider the spin Hamiltonian only in the 
two outer layers. Spin-orbit coupling in graphene induced
by TMDCs is of the valley-Zeeman type \cite{Gmitra2015}, meaning that $\lambda^{A_l}_I \approx 
-\lambda^{B_l}_I$. 

The numerical values for the parameters of the Hamiltonian 
$\hat{h}(\textbf{k},\tau) = \hat{h}_0 + \hat{h}_{\text{prox}}$ 
are taken from the \emph{ab initio} results of Ref.~\cite{PhysRevB.105.115126}. The calculated
low-energy band dispersions for pristine RTG, MoSe$_2$/RTG/WSe$_2$, and CGT/RTG/CGT with both parallel and antiparallel CGT magnetizations. 
are shown in Fig.~\ref{fig:scheme}(d).

We describe the correlation effects in RTG by Hamiltonian $\hat{H}=\hat{H}_{\text{kin}}+\hat{H}_{\text{int}}$,
\begin{align}
    \hat{H}_{\text{kin}}&=\sum_{\textbf{k}\tau,si,s'j}\left[\hat{h}(\textbf{k},\tau) - \mu \hat{I}\right]_{si,s'j} \hat{c}^{\dagger}_{s\tau i}(\textbf{k})\hat{c}_{s^\prime\tau j}(\textbf{k}),
    \label{eq:int}\\
    \hat{H}_{\text{int}}&=U\left(n_{\uparrow K}\,n_{\downarrow K}+n_{\uparrow K^{\prime}}\,n_{\downarrow K^{\prime}}\right)+
    Vn_{K}\,n_{K^{\prime}},
    \label{eq:int2}
\end{align} 
where $\mu$ stands for the chemical potential and $\hat{c}^{(\dagger)}_{s\tau i}(\textbf{k})$ 
is the annihilation (creation) operator for a Bloch electron with spin $s=\uparrow\hspace{-1.22mm}/\hspace{-1.22mm} \downarrow$ in valley $\tau={K/K^\prime}$ on RTG sublattice $i$ with a valley-momentum $\textbf{k}$.
The intra- and intervalley density interactions are described by repulsive (positive) couplings $U$ and $V$, respectively, while $n_{s\tau}=\sum_{|\textbf{k}|<\Lambda}\sum_{i}\hat{c}^{\dagger}_{s\tau i}(\textbf{k})\hat{c}^{\phantom{\dagger}}_{s\tau i}(\textbf{k})$ stands for the spin-valley number operator in the $s\tau$-channel cut-off by momentum $\Lambda$;
$n_{\tau}=n_{\uparrow\tau}+n_{\downarrow\tau}$ stands for the valley-resolved number operator. 
Specifically, we consider SU(4)-symmetric interactions by setting $U = V= 19$~eV~\cite{You2019, PhysRevB.105.134524}, and $\Lambda=0.06~\AA^{-1}$, yielding the pristine RTG phase diagram consistent with experimental findings~\cite{Zhou2021_1, Zhou2021_2}. We use the same interaction parameters for the encapsulated cases as well. 

\paragraph{Methodology.} To systematically determine the correlated phase diagram of  our model of proximitized RTG we employ the RPA. We first calculate the static irreducible susceptibility $\chi^0$ in spin-valley indices by employing the generalized Lindhard's susceptibility~\cite{PhysRevLett.101.087004, Graser2009, PhysRevB.83.100515,supplemental} at 4.2~K using the non-interacting Hamiltonian $\hat{H}_{\text{kin}}$, Eq.~(\ref{eq:int}). The RPA-corrected susceptibility equals $\chi=\left[1-\chi^0\Gamma\right]^{-1}\chi^0$, where $\Gamma$ is the fully-irreducible vertex function~\cite{supplemental}. 
A correlated phase emerges if the highest eigenvalue $\lambda_c$ of $\chi^0\Gamma$ becomes greater or equal to unity. Correspondingly, $\chi$ diverges, and the system becomes unstable.

We find the possible correlated phases $\hat{\Phi}$ by diagonalizing 
$\chi^0 \Gamma$ at $\mu = 0$ and $u_d = 0$. Then, varying the doping and displacement field, we compute the corresponding $\chi^0$. 
For each $\hat{\Phi}$ we estimate the critical parameter $\lambda^{\hat{\Phi}}_c=\langle\hat{\Phi}|\chi^0 \Gamma|\hat{\Phi}\rangle/{\lVert \hat{\Phi} \rVert^2}$. The dominant instability
at given $\mu$ and $u_d$ is determined by maximal $\lambda^{\hat{\Phi}}_c$. Only the states $\hat{\Phi}$ listed in Table~\ref{tab} are realized as
dominant in our models of MoSe$_2$/RTG/WSe$_2$ and CGT/RTG/CGT heterostructures.

\begin{figure}
    \centering
    \includegraphics[width=.5\textwidth]{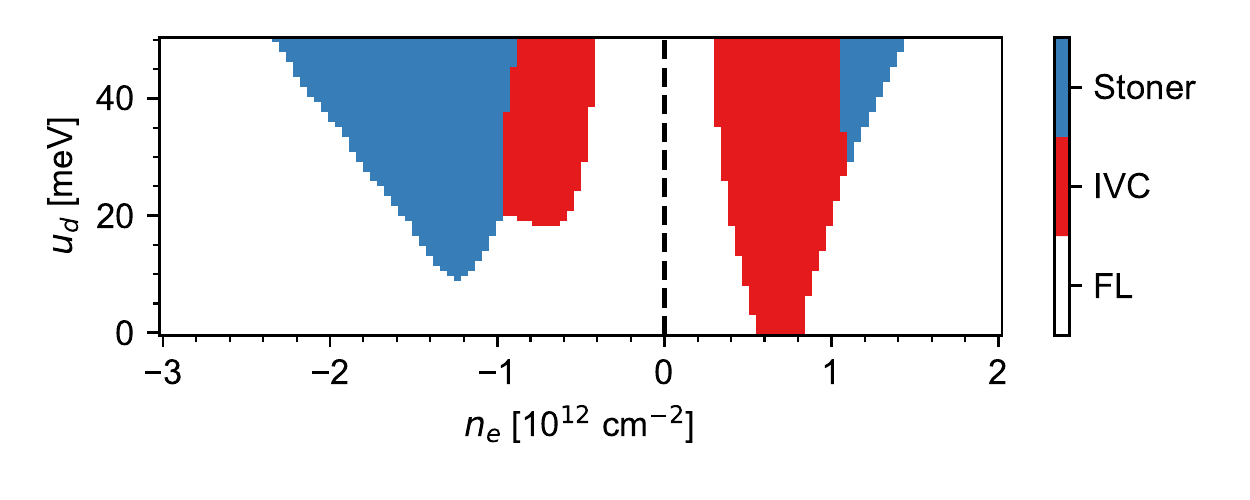}
    \caption{Calculated phase diagram of pristine RTG for varying doping $n_e$ and displacement field $u_d$. 
    There are two dominant phases: the intervalley coherent phase (IVC) and Stoner instability. 
    The white background corresponds to a stable Fermi liquid (FL). }
    \label{fig:free}
\end{figure}


We first explore the correlated phase diagram of pristine RTG, shown in Fig.~\ref{fig:free}, featuring IVC and the Stoner instabilities, as also reported earlier~\cite{PhysRevB.105.134524, Chatterjee2022}. While the Stoner instability is local, an IVC state describes a spatially
modulated phase by wave vector $\mathbf{q}$ which connects $K$ and $K^\prime$ valleys. 
The IVC and Stoner phases exhibit some degeneracies, as listed in Table I, due to the high symmetry of pristine RTG. 

\begin{figure}
    \centering
    \includegraphics[width=.5\textwidth]{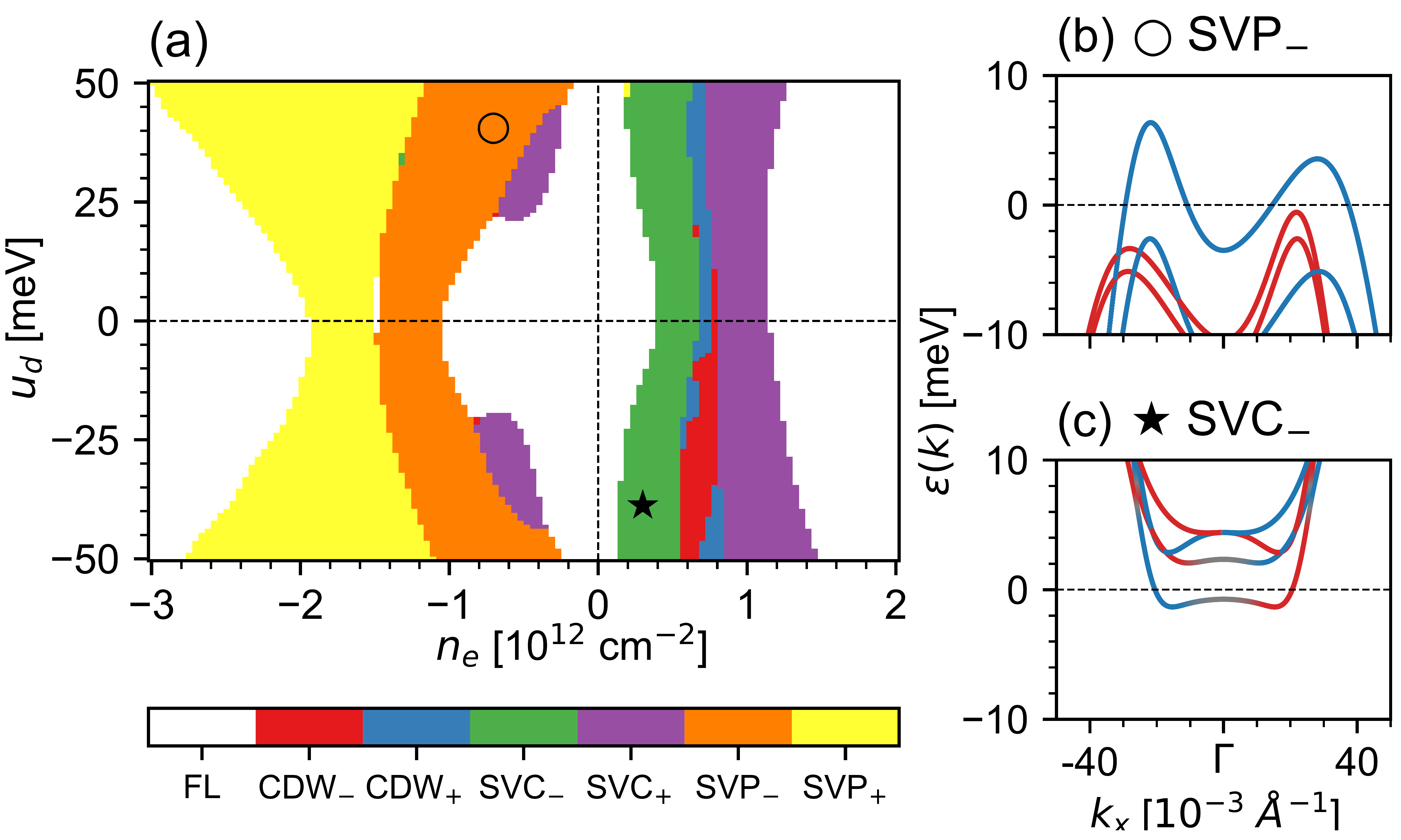}
    \caption{
    (a) Calculated phase diagram of a model MoSe$_2$/RTG/WSe$_2$ heterostructure in the space of displacement field $u_d$ and doping density $n_e$. 
    Six symmetry-breaking phases: CDW$_{\pm}$ (charge density wave), SVC$_{\pm}$ (spin-valley coherence), and SVP$_{\pm}$ (spin-valley polarized state), marked by different colors, are predicted to occur if the Fermi liquid state (white) is unstable.    
    Hartree-Fock excitation dispersions folded to the $\Gamma$ point for
    (b)~$u_d=40$~meV and $n_e=-0.7\cdot10^{12}$~cm$^{-2}$, and 
    (c)~$u_d=-40$~meV and $n_e=0.3\cdot10^{12}$~cm$^{-2}$. Red, blue, and gray lines mark
    spin-up, spin-down, and spin-unpolarized states. 
    The zero of energy corresponds to the 
    self-consistently computed Fermi level.
    }
    \label{fig:tmdc}
\end{figure}

\paragraph{Spin-orbit proximity effects.}
We now turn to a model MoSe$_2$/RTG/WSe$_2$ heterostructure ~\cite{PhysRevB.104.075126, PhysRevLett.125.196402} which exemplifies the SO effects on the correlated physics. 
While proximity SO coupling can be tuned by twisting~\cite{PhysRevB.104.195156, ZollnerTwist}, 
we consider here a generic set of SO parameters~\cite{PhysRevB.105.115126} which describe 
a zero twist-angle heterostructure. 

Fig.~\ref{fig:tmdc}(a) displays the obtained phase diagram. There are noticeable differences from the case of pristine RTG. In particular, the Stoner phase separates into two SVP$_{\pm}$ states, 
both spin-valley polarized along the spin-$z$ direction, as the single-particle valley-Zeeman coupling~\cite{Gmitra2015}.  
Also, the degeneracy of the IVC phase is removed, resulting in four distinct states which correspond to either a spin-valley coherent state, SVC$_{\pm}$, or a charge density wave, CDW$_{\pm}$, originating from specific spin flavors. 

The SVC$_{\pm}$ phases manifest at the electron doping near vHS split by SO fields. 
It can be explained by means of spin-valley correlations triggered by the valley-Zeeman SO coupling~\cite{Gmitra2015, supplemental}. The SVC$_{\pm}$ phases, $\hat{\Phi}_{\text{SVC}_{+}}=c^{\dagger}_{\downarrow K}c_{\uparrow K^{\prime}}+c^{\dagger}_{\uparrow K^{\prime}}c_{\downarrow K}$ and $\hat{\Phi}_{\text{SVC}_{-}}=c^{\dagger}_{\uparrow K}c_{\downarrow K^{\prime}}+c^{\dagger}_{\downarrow K^{\prime}}c_{\uparrow K}$, can be seen
as inter-valley spin-flipping hopping conserving the spin-valley quantum number. At the same time, CDW$_{\pm}$ phases, $\hat{\Phi}_{\text{CDW}_{+}}=c^{\dagger}_{\uparrow K}c_{\uparrow K^{\prime}}+c^{\dagger}_{\uparrow K^{\prime}}c_{\uparrow K}$ and $\hat{\Phi}_{\text{CDW}_{-}}=c^{\dagger}_{\downarrow K}c_{\downarrow K^{\prime}}+c^{\dagger}_{\downarrow K^{\prime}}c_{\downarrow K}$, appear between the split vHS due to the sign balance of the product of spin and valley quantum numbers on the Fermi surface. 
An almost degenerate nature of these phases is predicted, while their separation largely depends on the model parameters on the meV scale due to the insignificant difference in the critical parameter $\lambda^{\hat{\Phi}}_c$.

We performed self-consistent calculations of the correlated band structure of MoSe$_2$/RTG/WSe$_2$ using the Hartree-Fock method~\cite{supplemental}. Figures~\ref{fig:tmdc}~(b)~and~(c) show the emergent correlated bands as Hartree-Fock excitation dispersions folded into $\Gamma$ point, as would be the case for $3\times3$ unit cell. Panel~(b) represents the correlation-modified electronic band structure of the SVP$_{-}$ phase, while panel~(c) displays the corresponding band structure of the SVC$_{-}$ phase. For both, we calculated the Hartree-Fock self-energy $\hat{\Sigma}$ and based on it we estimated the correlated gaps, 
$\Delta_{\hat{\Phi}}=(\Tr{\hat{\Sigma}\hat{\Phi}})/{\lVert \hat{\Phi} \rVert} $, that are in magnitude comparable to the strength of the proximity-induced SO couplings, particularly,
$\Delta_{\text{SVP}_{-}}=3.935$~meV and $\Delta_{\text{SVC}_{-}}=2.770$~meV. 
Therefore, by adding an additional term $\hat{H}_{\Delta}=\Delta_{\hat{\Phi}}\hat{\Phi}/\lVert \hat{\Phi} \rVert$ into the single-particle Hamiltonian $\hat{H}_{\text{kin}}$, we can model an interacting system with symmetry broken phase on a self-consistent mean-field level. 
Particularly, the SVC$_{\pm}$ phase can be modelled by:
\begin{equation}
    \hat{H}_{\text{SVC}_{\pm}}=\frac{\Delta_{\text{SVC}_{\pm}}}{2\sqrt{2}}
    \sum_{i,s\tau,s'\tau'}
    \left[s_x\tau_x\pm s_y\tau_y\right]_{s\tau,s'\tau'} \hat{c}^{\dagger}_{s\tau i}(\textbf{k})\hat{c}_{s'\tau'i}(\textbf{k}),
\end{equation}
which has an effective single-particle dispersion describing spin-flip-valley-flip coupling, emerging due to the (intervalley) Coulomb interactions and the valley-Zeeman coupling. Such an interaction is expected to affect spin transport and spin dynamics profoundly. 

\begin{figure}
    \centering
    \includegraphics[width=.5\textwidth]{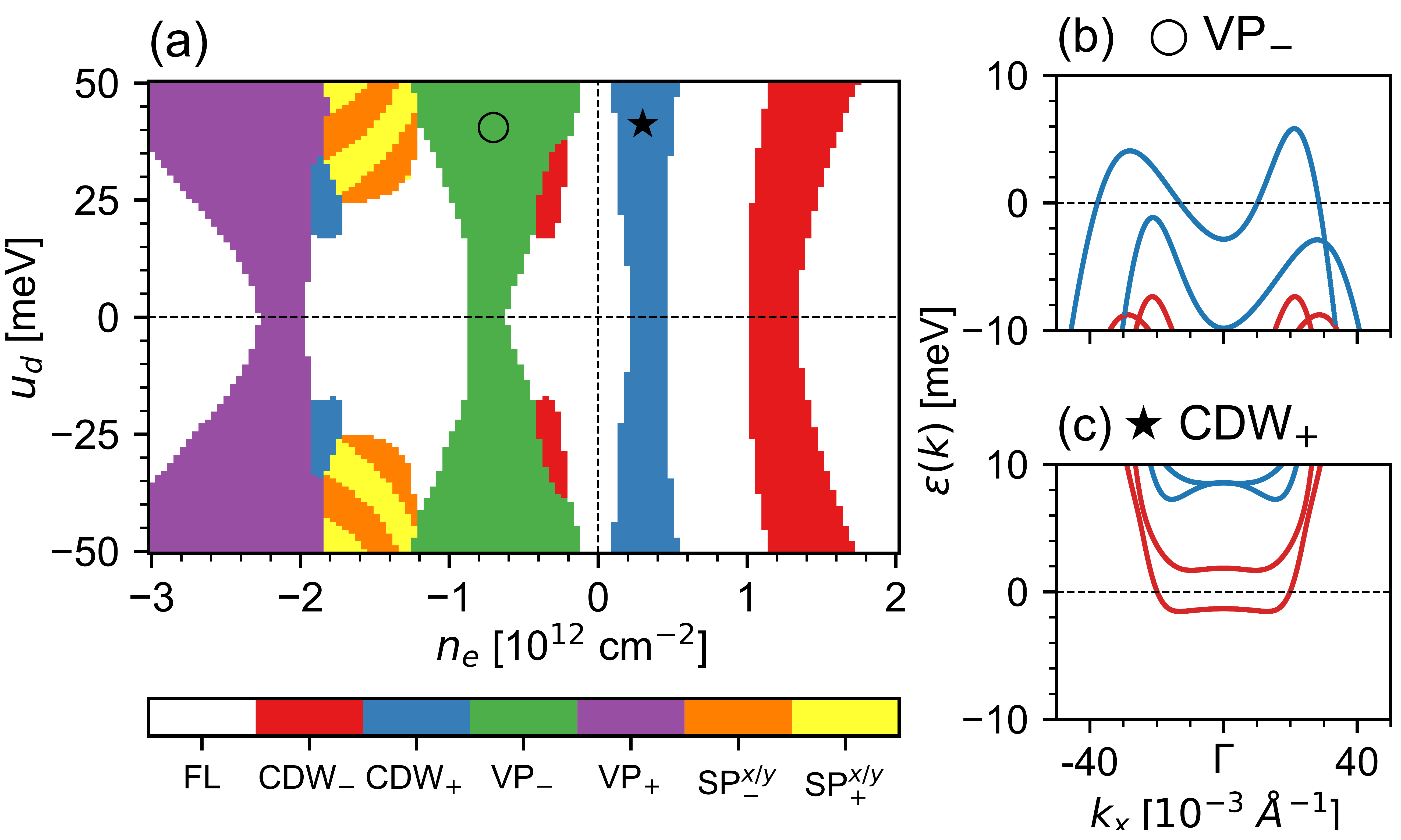}
    \caption{
    (a) Calculated phase diagram of CGT/RTG/CGT electron system with ferromagnetic CGT configuration 
    in a parameter space of displacement field $u_d$ and electron doping $n_e$. 
    Six symmetry broken phases, such as CDW$_{\pm}$ - charge density wave, VP$_{\pm}$ - valley-polarized state, and SP$^{x/y}_{\pm}$ - in-plane spin-polarized state are predicted and displayed by different colors, along with the Fermi liquid (FL) state.
    (b)~and~(c) Correlated band structures of the CGT/RTG/CGT folded to $\Gamma$ point for ferromagnetic CGT alignment at displacement field $u_d=40$~meV with two different electron dopings: (b)~$n_e=-0.7\cdot10^{12}$~cm$^{-2}$ and (c)~$n_e=0.3\cdot10^{12}$~cm$^{-2}$. 
    The color coding reflects spin expectation values, with red signifying spin-up, blue spin-down, and grey spin-unpolarized case. The zero of energy corresponds to the self-consistently determined Fermi level.
    }
    \label{fig:cgt-fm}
\end{figure}

\begin{figure}
    \centering
    \includegraphics[width=.5\textwidth]{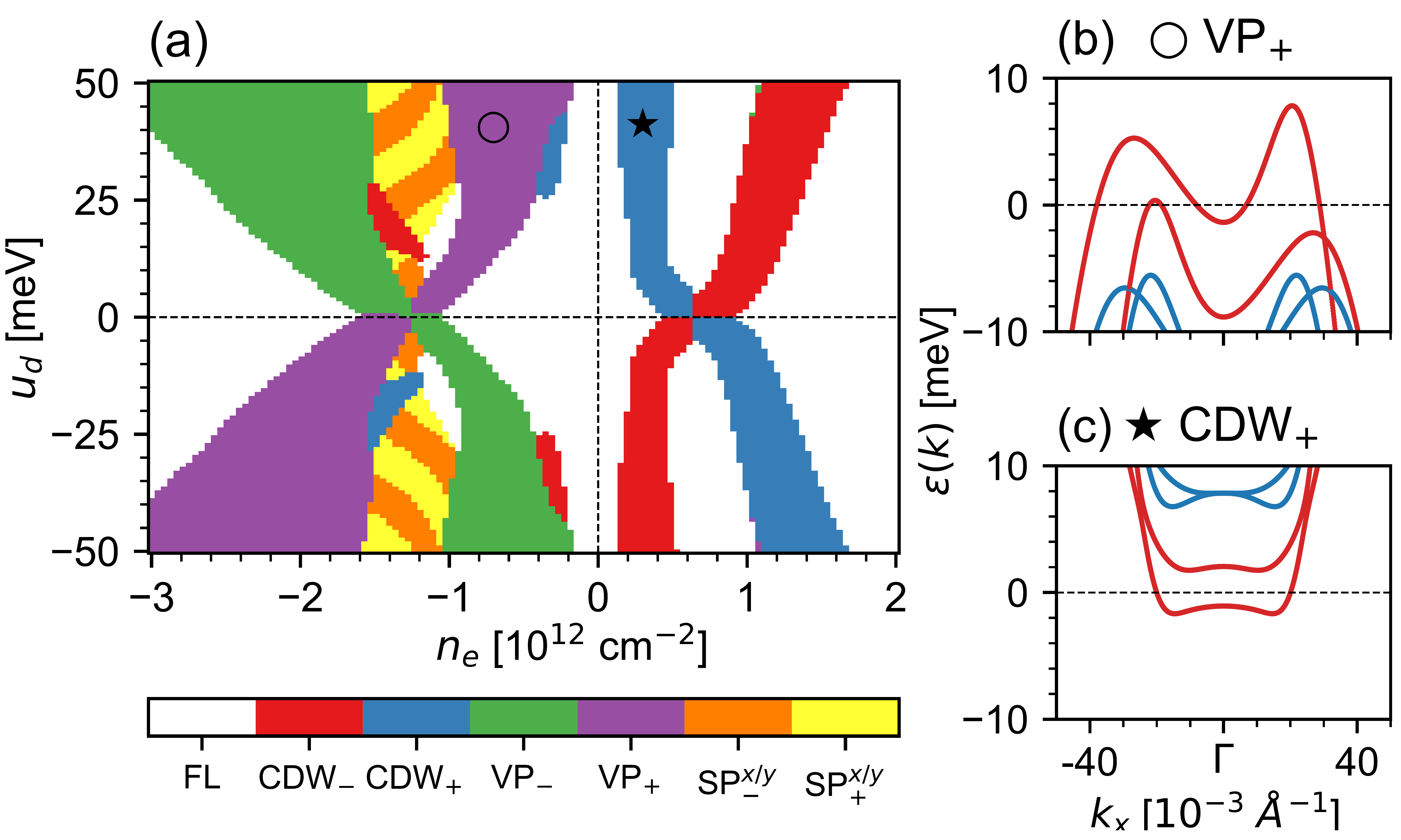}
    \caption{
    (a)~Phase diagram of the CGT/RTG/CGT electron system with antiferromagnetic CGT alignments in a parameter space of displacement field $u_d$ and electron doping $n_e$. Six phases with symmetry breaking, such as CDW$_{\pm}$ - charge density wave, VP$_{\pm}$ - valley-polarized state, and SP$^{x/y}_{\pm}$ - in-plane spin-polarized state is predicted along with the Fermi Liquid (FL) phase.
    (b)~and~(c) Correlated band structure of the CGT/RTG/CGT with antiferromagnetic CGT layers folded to $\Gamma$ point at displacement field $u_d=40$~meV and electron doping level (b)~$n_e=-0.7\cdot10^{12}$~cm$^{-2}$ and (c)~$n_e=0.3\cdot10^{12}$~cm$^{-2}$. The color coding scheme represents the spin expectation values, with red signifying spin-up, blue spin-down, and grey spin-unpolarized case. 
    The zero of energy corresponds to the self-consistently calculated Fermi level.
    }
    \label{fig:cgt-afm}
\end{figure}

\paragraph{Exchange proximity effects.}
Encapsulating RTG by ferromagnetic CGT induces spin-splitting of the RTG bands, depending on the relative magnetization orientation of the
CGT layer (parallel or antiparallel)~\cite{PhysRevB.105.115126}.
Also, here the twist-angle can affect the proximity effect. However, we are interested in what can happen in principle and adopt the zero-twist
heterostructure model parameters~\cite{ZollnerTwist}.

The calculated CGT/RTG/CGT phase diagram with the parallel CGT magnetizations is shown in Fig.~\ref{fig:cgt-fm}(a). 
The CDW$_{\pm}$ phase in the electron doping regime appears near vHS. The Stoner phase is manifested in valley-polarized VP$_{\pm}$ and spin-in-plane polarized SP$_{\pm}^{x/y}$ states. 
The absence of the SVC$_{\pm}$ and SVP$_{\pm}$ phases can be accounted for by the lack of the spin-valley coupled physics due to the absence of the valley-Zeeman SO coupling. Similar to the CDW$_{\pm}$ phase in MoSe$2$/RTG/WSe$2$, the SP$_{+}^{x/y}$ and SP$_{-}^{x/y}$ phases exhibit close degeneracy, and their splitting strongly depends on the model parameters.

We also examined the emergent electronic band structure of the CGT/RTG/CGT using the Hartree-Fock method.
Fig.~\ref{fig:cgt-fm}(b) shows the VP$_{-}$ phase, and Fig.~\ref{fig:cgt-fm}(c) presents the CDW$_{+}$ correlated band structure, both folded to $\Gamma$ point. Also, in this case, the correlated gaps match EX coupling amplitudes, particularly, $\Delta_{\text{VP}_{-}}=3.983$ meV and $\Delta_{\text{CDW}_{+}}=2.182$ meV.

Finally, we calculated the CGT/RTG/CGT phase diagram with antiparallel CGT magnetizations, shown in 
Fig.~\ref{fig:cgt-afm}(a). It exhibits strong tunability and displacement field asymmetry due to the contrasting magnetic order of the adjacent CGT layers. Similar to the ferromagnetic case, Stoner instability appears as VP$_{\pm}$ and  SP$_{\pm}^{x/y}$ states mainly in hole doping regime, while IVC as CDW$_{\pm}$ states in electron-doped one. Contrary to the ferromagnetic case, in CGT/RTG/CGT heterostructure with antiferromagnetic CGT layer ordering, the sign of displacement field is strongly determining the $\pm$-sign of the underlying correlated states. Figures 
\ref{fig:cgt-afm}~(b)~and~(c) display correlated folded Hartree-Fock electronic band structures for two specific correlated phases of antiferromagnetic CGT/RTG/CGT, VP$_{+}$ and CDW$_{+}$, respectively. Correlated gaps for these two cases are commensurate with EX coupling amplitudes and read $\Delta_{\text{VP}_{+}}=3.734$ meV and $\Delta_{\text{CDW}_{+}}=2.139$ meV.

\paragraph{Conclusion.} By performing realistic simulations of electronic correlations at the RPA level and by extracting Hartree-Fock excitation spectra for selected phases, we predict a strong interplay between proximity-induced SO and EX interactions and Coulomb-correlations furnished
by vHS of RTG. While we specifically consider MoSe$_2$/RTG/WSe$_2$ and CGT/RTG/CGT heterostructures, we expect our findings to be more generally applicable, opening up new possibilities for synergies between correlation physics and spintronics. The proposed emergence of spin-valley interactions facilitated by electronic correlations in the presence of SO coupling is particularly exciting, as it can enable new spintronic phenomena in van der Waals heterostructures.

\begin{acknowledgments}
This work was funded by the Deutsche Forschungsgemeinschaft (DFG, German Research Foundation) 
SPP 2244 (Project No. 443416183), SFB 1277 (Project-ID 314695032), by the European Union Horizon 2020 Research and Innovation Program under contract number 881603 (Graphene Flagship), and by FLAG-ERA project 2DSOTECH.
D.K.~acknowledges partial support from the IM\-PULZ project IM-2021-26---SUPERSPIN funded by the Slovak Academy of Sciences.
\end{acknowledgments}

\end{document}